\documentclass[12pt]{article}
\usepackage[utf8]{inputenc}
\usepackage[T1]{fontenc}
\usepackage[margin=1in]{geometry}
\usepackage{amsmath}
\usepackage{graphicx,psfrag,epsf}
\usepackage{enumerate}
\usepackage{natbib}
\usepackage{url}
\usepackage{multicol}
\usepackage{tikz}
\usetikzlibrary{shapes.geometric, arrows}
\usepackage{array}
\usepackage{hyperref}
\hypersetup{
    colorlinks=true,
    citecolor=black,
    linkcolor=black,
    filecolor=black,      
    urlcolor=black}

\def\bSig\mathbf{\Sigma}

\newcommand{\indep}{\rotatebox[origin=c]{90}{$\models$}}

\begin{document}

\title{\bf Estimating hypothetical estimands with \\ causal inference and missing data estimators \\ in a diabetes trial}

  \author{ Camila Olarte Parra \\
    Department of Medical Statistics, \\ London School of Hygiene and Tropical Medicine, UK \vspace{0.3cm}\\
    Rhian M. Daniel \\
    Division of Population Medicine, Cardiff University, UK \vspace{0.3cm}\\
    David Wright \\ 
    Data Science \& Artificial Intelligence, \\ BioPharmaceuticals R\&D, AstraZeneca, Cambridge, UK \vspace{0.3cm}\\
    Jonathan W. Bartlett \\
    Department of Medical Statistics, \\ 
    London School of Hygiene and Tropical Medicine, UK}
  \maketitle

\begin{abstract}
    The recently published ICH E9 addendum on estimands in clinical trials provides a framework for precisely defining the treatment effect that is to be estimated, but says little about estimation methods. Here we report analyses of a clinical trial in type 2 diabetes, targeting the effects of randomised treatment, handling rescue treatment and discontinuation of randomised treatment using the so-called hypothetical strategy. We show how this can be estimated using mixed models for repeated measures, multiple imputation, inverse probability of treatment weighting, G-formula and G-estimation. We describe their assumptions and practical details of their implementation using packages in R. We report the results of these analyses, broadly finding similar estimates and standard errors across the estimators. We discuss various considerations relevant when choosing an estimation approach, including computational time, how to handle missing data, whether to include post intercurrent event data in the analysis, whether and how to adjust for additional time-varying confounders, and whether and how to model different types of ICE separately.

    
\end{abstract}

\section{Introduction}
\label{s:intro}

Following the recently-published ICH E9 addendum, defining the target treatment effect of a clinical trial, also known as the `estimand', includes identifying intercurrent events and strategies to deal with them \citep{ICHE9Addendum}. Examples of intercurrent events (ICE) include treatment discontinuation, use of rescue medication, death prior to measuring the outcome of interest or any event that occurs after treatment initiation that either affects the interpretation of the outcome or prevents it from being observed. 

In diabetes trials, rescue medication in case of inadequate glycaemic control should be available for ethical reasons because of the deleterious effect of elevated glucose levels. In this setting, one option is to target the effect of treatment in a way that includes any effect that the addition of rescue medication may have on the outcome (e.g. HbA1c). According to the ICH E9 addendum, this would correspond to using a \emph{treatment policy} strategy to deal with this ICE. This strategy, however, leads to an estimand that may mask (or, less commonly, exaggerate) the effect of the study drug itself whenever there is differential use of rescue medication between treatment arms  \citep{Holzhauer2015DiabetesRescueMed}. In particular, if there are higher levels of rescue medication in the control arm compared to the active arm (and if the rescue medication is more effective than the control arm medication alone), the treatment policy estimand may understate the pharmacological benefits of the active treatment. Estimating the treatment effect in the (hypothetical) absence of rescue medication use can then be of interest for certain stakeholders. In this case, the use of rescue medication would be handled following a so-called \emph{hypothetical} strategy, targeting what would have been observed in the trial had rescue medication not been made available to patients.

In section \ref{s:example} a trial in type 2 diabetes patients is described as a motivating example. Although the trial did not specify an estimand, we believe the primary estimand would have used a hypothetical strategy to deal with rescue medication and treatment discontinuation given the method of statistical analysis. This estimand is the main focus of this paper and our aim is to illustrate how different estimation strategies can be applied in a real life scenario. Missing data methods are typically used to estimate such estimands, because the hypothetical outcome values that would have ensued in the absence of the ICE are missing. These include mixed-model repeated measures (MMRM) and multiple imputation (MI). Causal inference estimators, like G-formula, inverse probability of treatment weighting (IPTW) and G-estimation, have thus far been rarely used in clinical trials, presumably because they were mostly developed in the context of observational rather than randomised studies. In earlier work, we showed how causal inference estimators can be used to estimate hypothetical estimands, with the potential for improved statistical efficiency over estimates obtained using missing data methods \citep{Olarte2021hypothetical}. Here, we demonstrate feasible ways to implement the different estimators using existing statistical packages and discuss how to tackle challenges encountered in real life settings including missing data. Applying the different estimators to a common dataset allows us to compare their properties in terms of statistical assumptions, efficiency, and computational time. 

The structure of this paper is as follows. In Section \ref{s:example}, we describe the diabetes trial to be analysed and in Section \ref{s:estimands}, we define our hypothetical estimand of interest and assumptions under which it can be identified. We outline step by step how to implement the different estimators to target this estimand in Section \ref{s:methods}, report the results in Section \ref{s:results} and close with discussion and recommendations in Section \ref{s:discussion}. 

\section{Motivating example}
\label{s:example}
We analysed data from a randomised controlled trial (RCT) where type 2 diabetes patients on metformin monotherapy who had inadequate glycaemic control were randomised at baseline to receive add-on dapagliflozin ($n=299$), dapagliflozin and saxagliptin ($n=305$), or glimepiride ($n=302$) \citep{Muller2018Diabetes}. Their HbA1c and fasting plasma glucose (FGP) were measured at baseline and then periodically to assess their response. During the first 3 months, visits  occurred every 2 weeks and then every 12 weeks up to 52 weeks. The main outcome of interest ($Y$) was change in HbA1c from baseline to the final visit after 52 weeks of follow-up. The treatment effects reported compared the outcomes of the patients randomised to dapagliflozin, and those randomised to dapagliflozin and saxagliptin, to those in the glimepiride group. 

Insulin was indicated as rescue medication for patients with inadequate glycaemic control. The study protocol specified different thresholds, based on FPG at earlier visits and HbA1c at later visits, for considering rescue medication at each visit. According to the study protocol, if patients exceeded the relevant threshold they then had an extra visit and their FPG or HbA1c value was measured again. Rescue was then initiated if this value exceeded the threshold. Once rescue medication was started, patients continued taking it for the remainder of the study. There were no deaths observed during the study period. 

For simplicity, we will focus on the comparison of dapagliflozin and saxagliptin to glimepiride. We chose this comparison to illustrate the potential effect of an imbalance in use of rescue medication, given that in the glimeperide arm 12.3\% received rescue medication while in the dapagliflozin and saxagliptin arm only 6.2\% received it. 

Access to this trial data was approved by the sponsor AstraZeneca and requested via the platform Vivli (Vivli Data Request: 6764). 

\section{Estimand and identifiability assumptions}
\label{s:estimands}
To precisely define our estimand of interest, we will introduce some notation. Let $A$ denote the randomised treatment and $E_k$ be an indicator for the occurrence of the ICE (use of rescue medication or discontinuation of treatment) at visit $k$. Patients who started rescue or discontinued treatment, remained on rescue or without study treatment for the rest of the follow up. The overbar denotes the history of a variable up to and including visit $k$ (e.g. $\overline{E}_k$) or throughout the entire follow-up (e.g. $\overline{E}$). $Y_k^{a,\overline{e}_k}$ denotes the potential change in HbA1c from baseline that would be measured at visit $k$ if we were to set treatment $A=a$ and ICE occurrence $\overline{E}_k=\overline{e}_k$. Our estimand of interest is then:

\begin{equation}
\label{eq:estimand1}
    \mathbf{E} \left(Y_{10}^{a=1,\overline{e}=\overline{0}}-Y_{10}^{a=0,\overline{e}=\overline{0}}\right)
\end{equation}

In words, this is the effect (as a mean difference) on the change in HbA1c measured at the final (10th) week 52 visit ($Y_{10}$) of dapagliflozin and saxagliptin ($A=1$) compared to glimeperide ($A=0$) as add-on medications, if rescue medication had not been made available and patients had continued taking their assigned treatment during the full follow-up ($\overline{E}=\overline{0}$) . 



\begin{figure}
     \centering
    \resizebox{0.8\textwidth}{!}{ 
    \begin{tikzpicture}
    \node (A) at (1.50,-2.00) {$A$};
    
    \node (L_0) at (0.00,0.00) {$L_0$};
    \node (L_1) at (4.00,0.00) {$L_1$};
    \node (L_2) at (8.00,0.00) {$L_2$};
    \node (L_3) at (12.00,0.00) {$L_3$};
    
    \node (Y_1) at (5.50,-2.00) {$Y_1$};
    \node (Y_2) at (9.50,-2.00) {$Y_2$};
    \node (Y_3) at (13.50,-2.00) {$Y_3$};
    
    \node (ICE_1) at (7.00,-4.00) {$E_1$};
    \node (ICE_2) at (11.00,-4.00) {$E_2$};
    
    \draw [->] (L_0) edge (L_1);
    \draw [->] (L_0) edge (Y_1);
    \draw [->] (L_0) edge (ICE_1);
    \draw [->] (L_0) to[out=20,in=170] (L_2);
    \draw [->] (L_0) edge (Y_2);
    \draw [->] (L_0) to[out=-30,in=-150] (ICE_2);
    \draw [->] (L_0) to[out=20,in=170] (L_3);
    \draw [->] (L_0) edge (Y_3);
    
    \draw [->] (L_1) edge (Y_1);
    \draw [->] (L_1) to[out=-30,in=120] (ICE_1);
    \draw [->] (L_1) edge (L_2);
    \draw [->] (L_1) edge (Y_2);
    \draw [->] (L_1) edge (ICE_2);
    \draw [->] (L_1) to[out=20,in=170] (L_3);
    \draw [->] (L_1) edge (Y_3);
    
    \draw [->] (L_2) edge (Y_2);
    \draw [->] (L_2) to[out=-30,in=120] (ICE_2);
    \draw [->] (L_2) edge (L_3);
    \draw [->] (L_2) edge (Y_3);
    
    \draw [->] (L_3) edge (Y_3);
    
    \draw [->] (A) edge (L_1);
    \draw [->] (A) edge (Y_1);
    \draw [->] (A) edge (ICE_1);
    \draw [->] (A) edge (L_2);
    \draw [->] (A) to[out=20,in=170] (Y_2);
    \draw [->] (A) edge (ICE_2);
    \draw [->] (A) edge (L_3);
    \draw [->] (A) to[out=20,in=170] (Y_3);
    
    \draw [->] (Y_1) edge (ICE_1);
    \draw [->] (Y_1) edge (L_2);
    \draw [->] (Y_1) edge (Y_2);
    \draw [->] (Y_1) edge (ICE_2);
    \draw [->] (Y_1) edge (L_3);
    \draw [->] (Y_1) to[out=20,in=170] (Y_3);
    
    \draw [->] (Y_2) edge (Y_3);
    \draw [->] (Y_2) edge (ICE_2);
    
    \draw [->] (ICE_1) edge (L_2);
    \draw [->] (ICE_1) edge (Y_2);
    \draw [->] (ICE_1) edge (ICE_2);
    \draw [->] (ICE_1) to[out=60,in=200] (L_3);
    \draw [->] (ICE_1) edge (Y_3);
    
    \draw [->] (ICE_2) edge (L_3);
    \draw [->] (ICE_2) edge (Y_3);
    
    \end{tikzpicture}}

    \caption{Directed Acyclic Graph relating randomised treatment ($A$), baseline ($L_0$) and time-varying covariates ($L_k$), occurrence of the ICE at each visit ($E_k$) and repeated measurements of the outcome ($Y_k$).}
    \label{fig:DAG}
\end{figure}
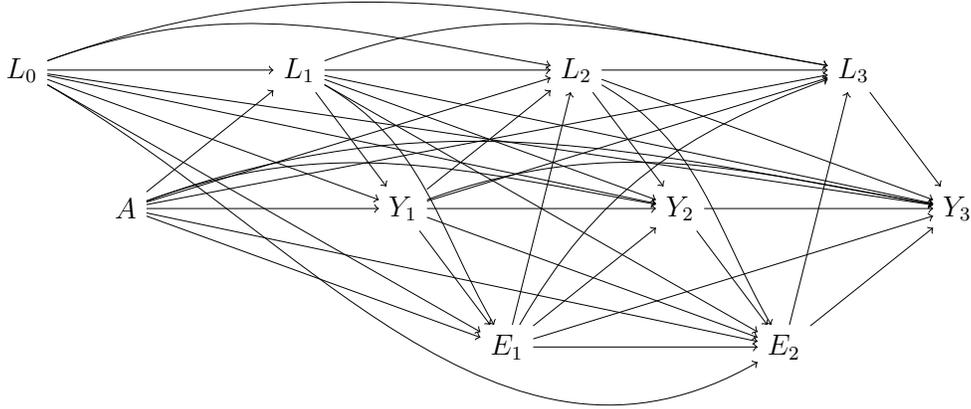

Figure \ref{fig:DAG} summarises the assumed causal structure between randomised treatment $A$, repeated measurements of outcome $Y$ and the occurence of the ICE in a simplified setting with 3 visits after baseline. $L_0$ denotes variables measured at baseline and $L_k$ variables measured at each follow-up visit $k$ . 

To estimate the mean potential outcome under each treatment, we make the following identifiability assumptions: 
\begin{enumerate}
    \item Consistency: for those randomised to arm $a$ who are free of the ICE throughout, their observed outcome is equal to their potential outcome $Y_{10}=Y_{10}^{a,\overline{e}=\overline{0}}$. 
    \item Sequential exchangeability: 
    together with $A$ and $\bar{E}_{k-1}$, $\bar{L}_k$ and $\bar{Y}_k$ are sufficient to control for confounding between each $E_k$ and $Y_{10}$. Formally, $Y_{10}^{a,\overline{e}=\overline{0}} \indep E_k | A=a, \bar{L}_k, \bar{Y}_k, \bar{E}_{k-1}=\bar{0}$ for all $k$ and for $a=0,1$.
 
    \item Positivity: if a participant is ICE-free up to visit $k-1$, there is a positive probability of again \emph{not} having the ICE at visit $k$, conditional on any possible combination of covariate (and past outcome) history and treatment arm: $P(E_k=0|A=a, \overline{L}_k = \overline{l}_k, \overline{Y}_k = \overline{y}_k, \bar{E}_{k-1}=\bar{0})>0$ for all $k$, for $a=0,1$ and for any $(\overline{l}_k,\overline{y}_k)$ such that the conditional covariate (and past outcome) density given treatment arm and `no ICE' is bounded away from zero at $(\overline{l}_k,\overline{y}_k)$. Note that the corresponding assumption $P(E_k=1|A=a, \overline{L}_k = \overline{l}_k, \overline{Y}_k = \overline{y}_k, \bar{E}_{k-1}=\bar{0})>0$ is not required because we are only interested in ICE-free potential outcomes. 
\end{enumerate}

The baseline covariates $L_0$ were age, sex, body mass index (BMI), systolic blood pressure, duration of diabetes and C-peptide (indicator of the production of insulin). These were chosen for our analyses in this paper since they were the variables presented in the standard Table 1 in the original publication. Choosing relevant variables to satisfy the exchangeability assumption requires consultation with experts in the corresponding therapeutic area at protocol stage to ensure the collection of the required information. 

In the version of the dataset to which we had access, age was grouped into 5-year categories, except for the first category that included patients from 18 to 30 years ($n=10$). We used the mid-point of each category to create a variable that was treated as a continuous variable in these analyses. When doing the primary analysis of a trial this would not be necessary as the actual age would be available; this approximation is only to facilitate the analysis here. BMI was available only as 3 categories (normal weight, overweight and obesity) and thus was included as categorical. 

Time-varying covariates $L_k$ were FPG and kidney function (estimated glomerular filtration rate, eGFR) measured at baseline and subsequent visits. As we explained in Section \ref{s:example}, the use of rescue medication was indicated according to the values of FPG. We chose to include only the scheduled measurements of FPG and not the additional measurements taken in those patients who exceeded the threshold at the scheduled visit. This was to avoid possible violations of the positivity assumption, but comes at the price of possibly violating the sequential exchangeability assumption. The impact of the residual confounding that may result from omitting the additional measurements is thought to be largely mitigated by the fact that we are accounting for HbA1c, a reflection of glycaemic control over approximately 3 months. The benefit (in terms of more closely satisfying sequential exchangeability) of including the additional measurements was judged to be out-weighed by the increased risk of positivity violations, and so $L_k$ only included the scheduled FPG measurement.

We also accounted for kidney function via inclusion of eGFR because impairment can lead to the discontinuation of these medications. As with other baseline characteristics, including repeated measurements of FPG and eGFR makes the exchangeability assumption more plausible. Once again we stress that we chose the variables that were available for the analysis and that careful planning at protocol stage with advice from subject matter experts is required to identify relevant variables to be measured at baseline and at each visit before the trial starts to satisfy this assumption.

\section{Methods}
\label{s:methods}

In this section, we describe the different estimators used to target the hypothetical estimand defined in Section \ref{s:estimands} and the corresponding software used to implement them. We start by replicating the original analysis with MMRM and then provide alternatives that include MI, IPTW, G-formula and G-estimation. We discuss variations of these approaches that include other relevant variables besides the ones included in the original analysis. The different methods and implementations are numbered sequentially as they are being described in the following subsections. These numbers are then used to link the description of the method with their corresponding results in Section \ref{s:results}. 

The code used to run these analyses is available at (\url{https://github.com/colartep/hypothetical_estimand_diabetes_trial}).

\subsection{Methods using only values before the ICE}
We first describe methods that only make use of data up until the occurrence of the ICE.

\subsubsection{Mixed model repeated measures (MMRM)}

The first step was to replicate the main analysis of the original paper that 
was performed in the set of patients who received at least one dose of the treatment, had a baseline value of HbA1c and at least one follow-up measurement. In general, we should use the information of all patients that were included in the trial and here we are restricting our analysis to this subset to replicate what was done in the original paper. This same set of participants was used for all the analyses we conducted. Only measurements of the outcome (change in HbA1c) before rescue treatment or discontinuation were included. This means that for individuals who had either ICE, we are setting $Y$ to missing at visits after the ICE occurred, even if it was actually observed.

The MMRM approach (\emph{Method 1}) was implemented using the \texttt{mmrm} package in R \citep{mmrmpackage} as follows:

\begin{enumerate}
    \item The data were arranged in long format, where each row corresponds to a visit of a particular subject.
    \item The visit covariate was converted to a factor variable, as required by the package. 
    \item Using the \texttt{mmrm}  function, a model was fitted using the default restricted maximum likelihood, with change in HbA1c from baseline at each visit as the response variable with fixed effects of treatment, visit, treatment by visit interaction, baseline HbA1C, and baseline HbA1c by visit interaction as covariates. An unstructured residual covariance structure was assumed. 
\end{enumerate}

MMRMs maximise the observed data likelihood, and provide valid inferences provided the full data model is correctly specified and the missing data satisfy the missing at random (MAR) assumption. Assuming the only missing data were those HbA1c values after occurrence of ICE, the MAR assumption, which is closely related to the sequential exchangeability assumption \citep{Olarte2021hypothetical}, would hold if we have included all common causes of the ICE and the outcome (change in HbA1c) in the MMRM model as covariates i.e. the common causes $L$ in Figure \ref{fig:DAG}. Since by design initiation of rescue medication was dependent on FPG, which was not used as a covariate in the MMRM, this MAR assumption is likely violated. However, the extent to which it is violated may be mitigated by the high correlation between FPG and HbA1c \citep{Holzhauer2015DiabetesRescueMed}. An advantage of not utilising the outcome values after the occurrence of the ICE is that this avoids the need to model the effect of the ICE on the outcome.

\subsubsection{Multiple imputation (MI)}
Using the same data as in the MMRM analysis and assuming MAR, the potential outcomes under no ICE were imputed using MI. We do not expect these results to be numerically identical to MRMM for reasons discussed by \citet{wang1998large} (including taking only a finite number of imputations); however, for a large sample size and a large number of imputations, the differences are expected to be very small.

To implement MI (\emph{Method 2}), we used the \texttt{mice} package in R \citep{Van2015micepackage}, as follows:

\begin{enumerate}
    \item The dataset was arranged in wide format where each row corresponds to a subject and measurements of change in HbA1c from baseline at each visit were recorded in separate columns. This was to ensure that the imputation model accounted for correlated observations from the same person. Only values prior to the ICE occurrence were included for this analysis. 
    \item Using the \texttt{mice} function and setting the method to `norm', i.e.\ Bayesian Linear Regression, we created 100 completed datasets using the default 5 iterations. Change in HbA1c at each visit was thus imputed using a normal linear model with randomised treatment, baseline HbA1c and HbA1c at all the other visits as covariates. This is equivalent to imputation from the joint multivariate model assumed in the MMRM model \citep{hughes2014joint}.
    \item Using each imputed dataset, we fitted a linear regression with change of HbA1c at the last visit as the outcome and baseline HbA1c and treatment group as covariates.
    \item The results were pooled across the imputed datasets following Rubin's rule using the \texttt{pool} function to obtain the corresponding standard errors. 
\end{enumerate}

An advantage of MI is that it allows the inclusion of a different set of covariates in the imputation and in the analysis models. By including additional covariates in the imputation model, the MAR assumption can be rendered more plausible. Thus, we also implemented other versions of MI that included the baseline (\emph{Method 3}) and time-varying variables (\emph{Method 4}) listed in section \ref{s:estimands} as covariates (without interactions) in the imputation models (Step 2).  

\subsubsection{Inverse probability of treatment weighting (IPTW)}

Next we applied inverse probability of treatment weighting (IPTW).
For this setting, the time-varying `treatment' corresponds to the the occurrence of the ICE. The weights for IPTW can be estimated using the \texttt{ipwtm} function of the \texttt{ipw} R package \citep{Van2011ipwpackage}. An important limitation to highlight is that the \texttt{ipw} package does not allow for missing values. To overcome this, we imputed the missing values before applying IPTW, as follows:

\begin{enumerate}
    \item The dataset was arranged in wide format. As with MMRM and MI, only values prior to the ICE were included. 
    \item Using the \texttt{mice} package, we created 100 completed datasets in the same way as with MI - step 2. 
    \item An ICE indicator was then constructed for each person at each visit, set to 1 if the patient had received rescue medication and/or discontinued treatment by that visit, and 0 otherwise.
    \item Each imputed dataset was rearranged to long format as is required by the \texttt{ipw} package. 
    \item Within each imputed dataset, we estimated the weights using the \texttt{ipwtm} function. This function fits a single pooled logistic regression model across all the observations. We included randomised treatment and HbA1c at the current visit, the previous visit and the average up until the previous visit as covariates. The function then estimates the weight for each patient at each visit, which is equal to the estimated probability of having the ICE, or of not having the ICE, whichever was observed. The final weight per patient corresponds to the product of their weights across all visits, for those who remain ICE-free, and up until the occurrence of the ICE, for those who had an ICE.  
    \item A weighted linear regression among those who remain ICE free throughout the entire follow up was fitted regressing the change in HbA1c at the final visit on treatment group and baseline HbA1c. Intuitively, the weighting gives particular emphasis to those patients who did not have the ICE but whose characteristics were very similar to those who did, to counter the fact that those who remain ICE free are a selected subset of the whole sample. 
    \item Estimates of the no-ICE potential outcome mean and treatment effect were obtained from the model. 
    \item The corresponding standard errors (SEs) were estimated via the non-parametric bootstrap using the \texttt{boot} package \citep{Canty2002bootpackage}. In order to reflect the estimation of the weights in these SEs, both the weights model (Step 4) and the weighted linear regression (Step 5) were included, i.e.\ the weights model and weighted linear regression were fitted on each bootstrap sample in turn. 
    \item  Estimation (including SE estimation) of the mean potential outcome and treatment effect was combined across the imputed datasets following Rubin's rules by applying the \texttt{pool} function and using the bootstrap variance estimate for the within-imputation variance \citep{Leyrat2019PSandMI}.

\end{enumerate}

As with MI, we have the flexibility to include additional variables to the ones included in the analysis model. These additional covariates, listed in Section \ref{s:estimands}, can be included in the model for the weights (step 3) and are beneficial if including them makes the sequential exchangeability assumption more plausible. We conducted IPTW without additional covariates besides baseline and repeated measurements of HbA1c and randomised treatment (\emph{Method 5}), another one also including additional baseline covariates (\emph{Method 6}) and finally one where both baseline and time-varying variables were included as covariates without interactions (\emph{Method 7}).

It is worth highlighting that in MMRM or MI there is no distinction between different ICEs. In contrast, for IPTW, as the ways covariates predict the occurrence of the two different ICE types may differ, an approach that distinguishes between the different ICE may be preferable. We could consider having separate logistic models for each ICE and then using the product of the probabilities of having each ICE to construct the weights. However this would imply that the events are independent. Alternatively, we can construct the ICE variable in step 3 as a factor variable to indicate whether no ICE occurred ($ICE=0$), only rescue ($ICE=1$), only discontinuation ($ICE=2$) or both occurred ($ICE=3$) and use this indicator to estimate the weights specifying, say, a multinomial logistic regression model in the argument family of the \texttt{ipwtm} function with all baseline and time-varying variables as covariates (\emph{Method 8}). 

For this and the following methods that required the bootstrap, we chose to draw 100 bootstrap samples, unless otherwise specified. Often one would use a larger number of bootstrap samples to minimize the Monte-Carlo error in the bootstrap estimate of variance. We chose a relatively small value here because the bootstrap variance estimates are themselves averaged across the 100 imputed datasets to calculate the average within-imputation variance. 

Estimating the effect on each imputed dataset and bootstrapping within each imputed dataset to obtain a within-imputation variance estimate is computationally faster than bootstrapping the observed dataset with missing values and then imputing in each bootstrap sample \citep{Schomaker2018Bootstrap}. Moreover, Rubin's variance estimator using bootstrapping to estimate the within-imputation variance has previously been shown to work well when combining MI with inverse probability weighting \citep{Leyrat2019PSandMI}.

\subsection{Methods exploiting post-ICE values}
To an increasing extent, clinical trials continue to collect information on patients after experiencing intercurrent events such as rescue medication and discontinuation of randomised treatment. We now describe methods that can exploit such information, potentially increasing statistical efficiency, but at the expense of having to make additional modelling assumptions.

\subsubsection{Parametric G-formula}
The G-formula by default makes use of measurements taken after the ICE occurs, in contrast to the approaches described previously. It is implemented in 
R in the \texttt{gfoRmula}  package \citep{mcgrath2020gformulapackage}. To perform G-formula (\emph{Method 9}), we followed these steps:

\begin{enumerate}
    \item The dataset was arranged in long format, with one row per visit per patient, and columns for each time-varying confounder and the outcome. In contrast to the previous methods described, any values that were observed after either ICE were included in this analysis. 
    As with IPTW, an ICE indicator variable was constructed to indicate whether that patient was using rescue medication or had discontinued randomized treatment by the corresponding visit. As in the \texttt{ipw} package, the \texttt{gfoRmula} package uses pooled models across visits. Thus, we created an additional variable to account for the time since the ICE occurred. This additional variable indicated the weeks since the ICE started, with this variable set to 0 if no ICE had occurred up until that visit or if it just occurred at the current visit. We chose this, as opposed to the number of visits since the ICE first occurred, because the time interval between visits varied, as explained in Section \ref{s:example}. From the perspective of the \texttt{gfoRmula} package, our `treatments' were the randomised treatment indicator, the ICE indicator and the weeks since ICE variable. We were thus simulating the potential outcomes under $\overline{a}_k=\overline{1}$, $\overline{e}_k=\overline{0}$, and $\overline{w}_k=\overline{0}$, and then under $\overline{a}_k=\overline{0}$, $\overline{e}_k=\overline{0}$ and $\overline{w}_k=\overline{0}$, where $w_k$ denotes the weeks since the ICE occurred at visit $k$. 

    \item Using the G-formula function for continuous outcomes from the \texttt{gfoRmula} package, we simulated the potential outcome under each treatment and no ICE for each individual. Note that here the outcome variable ($Y_{10}$) should be a separate column than intermediate measurements of $Y$ because, in the package, only the values at the last time point are used. We then need to specify a model for this time-varying covariate to use the function. This model for change in HbA1c from baseline included the randomised treatment, ICE indicator, the $w_k$ ICE variable, baseline HbA1c, HbA1c at the previous visit and an average HbA1c up to the previous visit. The same model specification was used for the $Y_{10}$ outcome model i.e. the model for change in HbA1c from baseline to the last visit. Using the arguments \texttt{histories} and \texttt{histvars} enable to include the lagged HbA1c i.e. measurement at the previous visit and the lagged average HbA1c without needing to create these additional variables manually. We set the argument \texttt{sim\_data\_b=TRUE}, so that the simulated dataset corresponding to the potential outcomes of interest were saved.
    \item The analysis model was a linear model regressing the simulated values of $Y_{10}$ on randomised treatment and baseline HbA1c. From this model, the mean potential outcome under each treatment and the treatment effect were obtained.
    \item The corresponding SEs were estimated via the non-parametric bootstrap using the \texttt{boot} package. As opposed to the other methods that use bootstrap, for G-formula we used 10,000 bootstraps because as we were not combining it with MI, we were not averaging across imputed datasets, and, as a result, a larger number of bootstrap was required. The bootstrap included both simulating the potential outcomes of interest (Step 3) and fitting the analysis model (Step 4).

\end{enumerate}

We also ran G-formula (\emph{Method 10}) including additional covariates in the HbA1c model (step 2) and including FPG and GFR as time-varying confounders. Besides randomised treatment, the ICE indicator, the $w_k$ ICE variable, current HbA1c, lagged HbA1c and lagged average HbA1c, the models for FPG, GFR and HbA1c included the same baseline and time-varying covariates for the previous methods listed in Section \ref{s:estimands}. We also included a lagged value and lagged average value of FPG and GFR in all the models. The final model using the simulated data (step 3) was not modified i.e. it only included randomised treatment and baseline HbA1c as covariates with simulated change in HbA1c at the last visit as response variable. As with IPTW, we also included an additional implementation where the ICE was a categorical variable indicating that the ICE did not occur ($ICE=0$), only rescue happened ($ICE=1$), only discontinuation happened ($ICE=2$) or both occurred ($ICE=3$) (\emph{Method 11}).

As opposed to the the \texttt{ipw} package, the \texttt{gfoRmula} package can be used with datasets with missing values. A single model is fitted for each time-varying confounder to the pooled long-form data. Any rows (measurements of a patient at a given visit) in which a missing value occurs in the response or covariates in these models are ignored by default by R's regression model fitting functions. The resulting `complete case' fits yield valid estimates provided missingness is independent of the response variable, given the covariates. In general, this assumption does not coincide with an MAR assumption, and indeed it may often be deemed more plausible than MAR \citep{white2010bias}.

\subsubsection{G-formula via MI}
A convenient alternative is to use a variant of standard MI to implement G-formula \citep{Bartlett2023GviaMI}. A potential advantage of this approach is that it also permits the imputation of both the missing actual and missing counterfactual data simultaneously.
The G-formula via MI approach (\emph{Method 12}) was implemented as follows: 

\begin{enumerate}
    \item  The dataset was in wide format i.e. one row per individual as with standard MI. As with G-formula, we included any observed values after the occurrence of the ICE.
    \item An ICE binary indicator was constructed for every visit and set to 1 if either or both ICE occurred by the given visit and 0 otherwise. As such, like before, once the indicator was 1, it remained 1 for the following visits.
    \item Using the \texttt{mice} function, we first imputed the missing actual data, creating 100 completed datasets. Change in HbA1c from baseline at each visit was imputed using a normal imputation model conditional on randomised treatment, HbA1c at the other visits, and the ICE indicators at each visit. 
    \item In each imputed dataset, we duplicated each row twice. One copy of each subject was assigned to treatment $A=1$ and the other to $A=0$. 
    \item In these duplicated rows, we set all the ICE indicators to 0 and all values of $Y_k$ to missing.
    \item The predictor matrix for the \texttt{mice} function, which specifies the covariates used in each conditional imputation model, was set to impute sequentially forwards in time i.e. imputing $Y_1$, using baseline values only,  imputing $Y_2$, using baseline values and values recorded at visit 1 and so forth (Figure \ref{fig:DAG}). 
    \item In each imputed dataset, we again used the \texttt{mice} function to impute the missing potential outcomes once. This second imputation step simulated values of $Y$ at each visit under no ICE in the duplicated rows.
    \item Using only the duplicated rows, a linear regression was fitted with $Y_{10}$ as the outcome and baseline HbA1c and randomized treatment as covariates. 
    \item The estimated mean outcome under each treatment and its difference were then averaged across all imputed datasets. As described in our previous work, Rubin's rules overestimates the variance for this setting, because the data used to fit the imputation models and the data used to fit the analysis model are not the same \citep{Bartlett2023GviaMI}. Thus, we use the variance estimator proposed by \citet{Raghunathan2003Varformula} when using MI to generate synthetic samples. The total variance is estimated as 
$T_{M} = (1+1/M) \hat{b}_{M} - \hat{v}_{M}$,  where  
$\hat{b}_{M}$ is the estimated between-variance and  $\hat{v}_{M}$ the estimated within-variance. For further details of this variance estimator for G-formula via MI, see \cite{Bartlett2023GviaMI}. 
\end{enumerate}

We also implemented a different version of G-formula via MI (\emph{method 13}) that included the set of baseline and time-varying covariates listed before for the previous methods in both imputation models (Step 3 and 7) and another one with all these covariates but with categorical indicators of the ICE (\emph{method 14}). 

\subsubsection{G-estimation}
G-estimation is an alternative approach that has recently been used for estimating hypothetical estimands \citep{Lasch2022G-estimation1,Lasch2022G-estimation2}. \citet{Loh2020gestimation} described how G-estimation can be used in mediation analysis to estimate controlled direct effects. For our setting, we can view the ICE as a mediator. As the hypothetical estimand targets the treatment effect in the absence of the ICE, it is equivalent to a controlled direct effect when setting the mediator to 0. We can use this procedure in combination with the bootstrap to obtain an estimate of its SE. As before, we dealt with missing data using MI first. 

We estimated the hypothetical estimand through G-estimation (\emph{Method 15}) as follows:
\begin{enumerate}
    \item The dataset was arranged in wide format. Any values after the ICE occurrence were included in this analysis.
    \item A binary indicator for the ICE was constructed for each visit in the same way as for G-formula via MI.
    \item Using the \texttt{mice} package, we first created 100 completed datasets. As with G-formula via MI, the imputation model included randomised treatment, repeated values of change in HbA1c from baseline and ICE binary indicators.   
    \item In each imputed dataset, we fitted a linear regression model by regressing change in HbA1c at the final visit ($Y_{10}$) on all the previous measurements of change in HbA1c as separate covariates, baseline HbA1c, randomised treatment and ICE indicators.  
    \item We then created a `modified' outcome variable for the previous visit, $R_{9}$, by subtracting the estimated effect of the ICE: $R_{9} = Y_{10} - \widehat{\psi}_{E_9} E_{9}$, where  $\widehat{\psi}_{E_9}$ is the estimated coefficient of the occurrence of the ICE in visit 9 $E_9$ taken from the model from step 4. 
    \item We then fitted a new linear regression model with $R_{9}$ as the outcome and all the previous measurements of HbA1c as separate covariates, as well as  randomised treatment and ICE indicators, saving the coefficient $\widehat{\psi}_{E_{8}}$ of $E_8$.
    \item We then repeated steps 5 and 6, creating $R_{k-1}=R_k-\widehat{\psi}_{E_{k-1}}E_{k-1}$, starting at $k=8$,  and regressing $R_{k-1}$ on the past, down to and including $k=1$.
    \item The treatment effect then corresponds to $\psi_{A}$ i.e. the coefficient of treatment in the model regressing $R_0$ on randomised treatment and baseline HbA1c.
    \item Finally we used non-parametric bootstrap (100 bootstrap samples) to estimate the SE. 
    \item We then pooled the estimates across the imputed datasets using Rubin's rules. 
\end{enumerate}

As with G-formula, we also implemented a version of G-estimation where we included the additional baseline and time-varying covariates in steps 1 and 2 and values up to $k$ in steps 4 and 6 (\emph{method 16}) and one where we have all the covariates and the categorical indicator of the ICE (\emph{method 17}). 

Table \ref{tab:estimators} shows a summary of the different estimators with the corresponding R package used, the data format required, how missing data were handled, whether it included post-ICE values and how the corresponding SEs were estimated. 

\begin{table*}
 \centering
 \def\~{\hphantom{0}}
  \caption{Summary of the different estimators and their implementation in R}
    \label{tab:baseline}
    \resizebox{\textwidth}{!}{
    \begin{tabular}{c|c|c|c|c|c}
    & & & Prior MI to  handle & & \\
    Method & R package & Data format & missing data & Post-ICE values & Standard error \\
    \hline
    MMRM & \texttt{mmrm} & Long & No & No & Likelihood-based \\
    MI & \texttt{mice} & Wide & Yes & No & Rubin's rules \\
    IPTW & \texttt{ipw} & Long & Yes & No & Bootstrap  \\
    G-formula & \texttt{gfoRmula} & Long & No & Yes& Bootstrap \\
    G-formula via MI & \texttt{mice} & Wide & Yes & Yes & Raghunathan's formula \\
    G-estimation & None & Wide & Yes & Yes & Bootstrap \\
    \end{tabular}}
\label{tab:estimators}
\end{table*}

\section{Results}
\label{s:results}

Table \ref{tab:baseline} summarises the characteristics of the patients randomised to each treatment arm of the trial. There are fewer patients than in the original trial publication because some of them withdrew consent ($n=33$). There were very few missing baseline values, with many variables having complete information and the rest having less than 0.5\% missing per variable. It is worth noting that most of the missing values belong to the dapagliflozin arm, that was not included in our analysis. Compared to the dapagliflozin + saxagliptin arm, there were more treatment discontinuations ($n=14$, 4.6\% vs.\ $n=7$, 2.3\%) and use of rescue medication ($n=37$, 12.3\% vs.\ $n=19$, 6.2\%) than in the glimeperide arm. 

\begin{table*}
 \centering
 \def\~{\hphantom{0}}
 \begin{minipage}{175mm}
  \caption{Patient characteristics}
    \label{tab:baseline}
    \resizebox{\textwidth}{!}{
    \begin{tabular}{l|cccc}
 & & 	Glimepiride & Dapagliflozin 10mg and
 &  \\
Variable
 & Dapagliflozin 10mg 
 & 1mg/2mg/4mg
 & Saxagliptin 5mg
 & Overall 
  \\

 &
(N=299) & 
(N=302) & 
(N=305) &
(N=906)  \\
\hline
Age (years), mean (SD) &	56.9 (9.59) &	58.2 (8.43) &	58.8 (7.98) &	58.0 (8.71) \\
Sex, n (\%) & & & &	\\			
\hspace{0.2cm} Women &	108 (36.1\%) &	98 (32.5\%) &	119 (39.0\%) &	325 (35.9\%) \\
\hspace{0.2cm} Men &	191 (63.9\%) &	204 (67.5\%) &	186 (61.0\%) &	581 (64.1\%) \\
Baseline Body Mass Index (kg/m $^2$)  & & & &	\\						
\hspace{0.2cm} 19 $<$ x $\leq$ 25 &	9 (3.0\%) &	9 (3.0\%) &	20 (6.6\%) &	38 (4.2\%) \\
\hspace{0.2cm} 25  $<$ x $\leq$ 30 &	76 (25.4\%)	& 83 (27.5\%) &	87 (28.5\%) &	246 (27.2\%) \\
\hspace{0.2cm} 30  $<$ x $\leq$ 80 &	212 (70.9\%) &	210 (69.5\%) &	198 (64.9\%) &	620 (68.4\%) \\
\hspace{0.2cm} Missing, n (\%) &	2 (0.7\%) &	0 (0\%)	& 0 (0\%) &	2 (0.2\%) \\
Baseline Systolic Blood Pressure (mmHg), mean (SD) &	138 (14.4) &	139 (13.0) &	139 (14.0) &	139 (13.8) \\
\hspace{0.2cm} Missing, n (\%) &	1 (0.3\%) &	0 (0\%)	& 0 (0\%) &	1 (0.1\%) \\
Baseline Waist Circumference (cm), mean (SD) &	111 (14.0) &	112 (13.2) &	109 (12.4) &	111 (13.2) \\ 
\hspace{0.2cm} Missing, n (\%) &	2 (0.7\%) &	3 (1.0\%) &	2 (0.7\%) &	7 (0.8\%) \\
Baseline Hip Circumference (cm), mean (SD) &	113 (12.5) &	112 (11.9) &	111 (11.4) &	112 (11.9) \\
\hspace{0.2cm} Missing, n (\%) &	4 (1.3\%) &	2 (0.7\%) &	3 (1.0\%) &	9 (1.0\%) \\
Years Since First Diagnose, mean (SD) &	6.88 (5.24) &	6.73 (5.14) &	7.39 (5.95) &	7.00 (5.46) \\
\hspace{0.2cm} Missing, n (\%) &	1 (0.3\%) &	0 (0\%)	& 0 (0\%) &	1 (0.1\%) \\
Baseline HbA1c (\%), mean (SD) &	8.29 (0.718) &	8.31 (0.753) &	8.25 (0.661) &	8.28 (0.711) \\
\hspace{0.2cm} Missing, n (\%) &	1 (0.3\%) &	3 (1.0\%) &	0 (0\%)	& 4 (0.4\%) \\
Baseline FPG (mmol/L), mean (SD) &	10.6 (2.31) &	10.4 (2.11) &	10.4 (1.99) &	10.5 (2.14) \\
Baseline eGFR (MDRD, mL/min/1.73m2), mean (SD) &	86.8 (18.8) &	85.8 (17.5) &	88.1 (19.7) &	86.9 (18.7) \\
Baseline C-Peptide (nmol/L), mean (SD) &	0.925 (0.356)	& 0.936 (0.345) &	0.920 (0.375) &	0.927 (0.359) \\
Discontinuation of randomised treatment, n (\%)	 & & & &	\\					
\hspace{0.2cm} No &	281 (94.0\%) &	288 (95.4\%) &	298 (97.7\%) &	867 (95.7\%) \\
\hspace{0.2cm} Yes &	18 (6.0\%)	& 14 (4.6\%) &	7 (2.3\%) &	39 (4.3\%) \\
Use of rescue medication, n (\%)  & & & &	\\						
\hspace{0.2cm} No	& 254 (84.9\%) &	265 (87.7\%) &	286 (93.8\%) &	805 (88.9\%) \\
\hspace{0.2cm} Yes &	45 (15.1\%) &	37 (12.3\%) &	19 (6.2\%) &	101 (11.1\%) \\

    \end{tabular}}
  
\end{minipage}
\vspace*{6pt}
\end{table*}

\begin{table*}
\caption{Number of missing HbA1c values by visit and treatment group}
    \begin{tabular}{l|c|c|c|c|c|c|c|c|c|c}
\hline
 &  \multicolumn{10}{c}{Visit} \\
   Treatment group  & 1 & 2 & 3 & 4 & 5 & 6 &7 &8 & 9 & 10 
     \\
     \hline
     
   Dapagliflozin and Saxagliptin & 5 & 2 & 5 & 7 & 5 & 10 & 7 & 7 & 14 & 4  \\
   Glimeperide & 9 & 8 & 8 & 12 & 10 & 14 & 11 & 15 & 16 & 14 \\
     \hline 
   \hspace{1cm} \emph{Setting post-ICE values to missing} \\
   Dapagliflozin and Saxagliptin & 9 & 9 & 9 & 17 & 14 & 14 & 19 & 27 & 33 & 29  \\
   Glimeperide & 15 & 16 & 15 & 22 & 16 & 19 & 29 & 49 & 58 & 55 \\
\hline

\end{tabular}
\label{tab:missing}
\vspace*{6pt}
\end{table*}

\begin{figure}
    \centering
    \resizebox{0.9\textwidth}{!}{
    \includegraphics{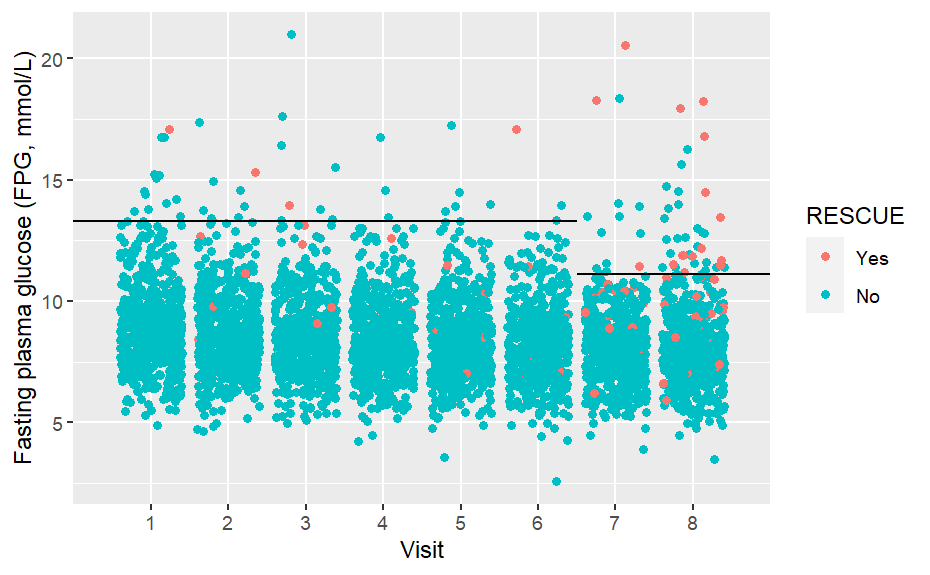}}
    \caption{Values of FPG per visit for all subjects. The black line corresponds to the threshold for indicating rescue at a given visit according to the protocol and the colour code indicates whether rescue was  initiated at that or an earlier visit.}
    \label{fig:positivity}
\end{figure}

Table \ref{tab:missing} summarises the numbers of missing outcomes per visit in each arm. At each visit, there were more missing values in the glimeperide arm than in the dapaglifloxin and saxagliptin arm. Treating post-ICE values as missing increases missing outcome values from 0.1-5\% to 3-20\% per visit. For example, in visit 9, which has the higher number of missing values, the missingness increases from $5\%$ to $15\%$ when outcome values after the ICE are deleted. This suggests that exploiting post-ICE values may increase efficiency. 

In Section \ref{s:example}, we noted that rescue medication at visits 1--8 was indicated according to the values of two measurements of FPG, only the first of which is used in our analyses. The omission of the second measurement is expected to avoid positivity violations, although near-violations are still a concern. Figure \ref{fig:positivity} shows a plot of the (first) FPG values per visit with the colour indicating whether rescue was initiated either at that visit, or an earlier visit. The threshold (which changed between visits 6 and 7) used for the FPG measurement is indicated by the black line, and the distribution of the red points both above and below the line reassures us that positivity is not violated, at least for the FPG variable, where the concern is greatest. 

The results for the different estimators are presented in Table \ref{tab:estimand1}. The estimates for the potential no-ICE outcome under dapagliflozin + saxagliptin are consistent across all the different variations of the methods, except for G-formula 10 and 11 that are somewhat larger in magnitude. The estimates of the mean potential no-ICE outcome under glimeperide have a greater variability across methods, compared to the variability of the estimates for dapagliflozin + saxagliptin, with MMRM, G-formula via MI and G-estimation yielding estimates of lower magnitude compared to MI and IPTW. Here, G-formula 10 and 11 also gave estimates of larger magnitude than the rest of the methods. The estimates of the treatment effect range from $-0.071$ to $-0.189$ across the different methods. Compared to the treatment effect estimate from the original published analysis, these are consistently slightly smaller in magnitude, which could be explained by the fact that the dataset contained 33 fewer patients, as noted earlier. 

For most methods, including additional covariates did not have much impact on the estimates, either of the mean potential outcomes or the treatment effect. This would suggest that the impact of additional confounding due to these covariates was small. For the G-formula, however there does seem to be an impact of including these additional covariates, particularly for the estimates of the two mean potential outcomes. As it was only with this method, it is unlikely due to confounding, and more likely either due to the modelling choices made for the pooled models used in G-formula, or the different assumptions made by G-formula regarding missing data, as described above.

Comparing methods that only use values before the ICE, IPTW had slightly larger SE as expected \citep{Carpenter2006Comparison}. Including values post-ICE has the potential to improve the efficiency of estimates \citep{Olarte2021hypothetical}, but in these analyses the SEs of G-formula, G-formula via MI and G-estimation were comparable to those of the estimators which only used data up until the ICE occurred. Allowing for distinct mechanisms for the two different ICEs in IPTW increased the SE, as one would expect. For the other methods, allowing for separate ICE mechanisms had relatively little impact on inferences.

In terms of computational time, MMRM and MI were much faster than the other methods. This is because they handle intermittent missing data in the same process as estimating potential outcomes and they do not require bootstrapping to obtain estimates of the SE. IPTW and G-estimation were much slower because they required handling missing data with MI as a first step and also bootstrapping. Somewhat unexpectedly, the slowest method was G-formula even though we did not combine it with MI to handle missing data. As explained before, the package fits a pooled model across visits so there are less models and parameters to be estimated compared to those methods fitting separate models per visit, such as MI or G-estimation. The computational time was greatly improved when using the alternative G-formula via MI that avoids the need for bootstrapping.  

\begin{table*}
 \centering
 \def\~{\hphantom{0}}
 \begin{minipage}{\textwidth}
  \caption{Potential outcome mean and treatment effect estimates under no rescue or discontinuation of randomized treatment using the different estimators}
    \label{tab:estimand1}
    \resizebox{\textwidth}{!}{
  \begin{tabular}{clcccr}
    No & Method &
Dapagliflozin and &
 & Treatment & Computational\\
&  &
Saxagliptin,  & Glimeperide,
 & effect, & time (mins) \\
&
& estimate (SE)  & estimate (SE) 
 & estimate (SE) \\
\hline
& \multicolumn{5}{c}{Using only values before the ICE} \\
\hline
& \multicolumn{1}{c}{MMRM}  \\
& Original published analysis  & -1.20 (0.05) &	-0.99 (0.05) & -0.21 (0.07)  \\
1 & Replicating original analysis &	-1.230 (0.0430) &	-1.093 (0.0449) & -0.137 (0.0621) & 1 \\
\hline

& \multicolumn{1}{c}{Multiple Imputation} \\
2 & HbA1c and treatment &	-1.234 (0.0425) &	-1.115 (0.0443) & -0.119 (0.0615) & 1 \\

3 & + all baseline covariates &	-1.234 (0.0421) &	-1.115 (0.0445) & -0.119 (0.0611) 
 & 1 \\
4 & + all time-varying covariates	& -1.234 (0.0420) &	-1.120 (0.0439) & -0.114 (0.0604) & 4 \\
\hline
& \multicolumn{1}{c}{IPTW*} \\ 
5 & HbA1c and treatment &	-1.249 (0.0430) &	-1.144 (0.0462) & -0.105 (0.0630) & 19\\
6 & + all baseline covariates &	-1.247 (0.0437) &	-1.143 (0.0468) & -0.104 (0.0638) & 32 \\

7 & + all time-varying covariates &	-1.215 (0.0521) &	-1.133 (0.0488) & -0.082 (0.0717) & 35 \\
8 & Separate ICE mechanisms & -1.207 (0.0611) & -1.136 (0.0493) & -0.071 (0.0783) & 368 \\
\hline
& \multicolumn{5}{c}{Exploiting post-ICE values} \\
\hline

& \multicolumn{1}{c}{G-formula} \\
9 & HbA1c and treatment &	-1.272 (0.0412) &	-1.125 (0.0453) & -0.147 (0.0600)  & 241 \\
10 & + all covariates &	-1.433 (0.0433) &	-1.246 (0.0465) & -0.187 (0.0627) & 1,506 \\

11 & Separate ICE mechanisms & -1.420 (0.0446) &	-1.231 (0.0477) & -0.189 (0.0629) & 1,561 \\

\hline
& \multicolumn{1}{c}{G-formula via MI} \\
12 & HbA1c and treatment &	-1.205 (0.0457) &	-1.062 (0.0533) & -0.143 (0.0637) & 2\\
13 & + all covariates  &	-1.238 (0.0501) &	-1.083 (0.0449) & -0.155 (0.0690) & 6 \\
14 & Separate ICE mechanisms &   -1.225 (0.0368) &	-1.070 (0.0469) & -0.155 (0.0611) & 6 \\
\hline
& \multicolumn{1}{c}{G-estimation*} \\
15 & HbA1c and treatment &	-1.212 (0.0440) &	-1.065 (0.0531) & -0.147 (0.0639) & 15 \\
16 & + all covariates & -1.211 (0.0446) &	-1.064 (0.0532) & -0.147 (0.0651) &  21 \\
17 & Separate ICE mechanisms & -1.211 (0.0446) & -1.064 (0.0532) & -0.147 (0.0651) &  22 \\
\hline
    \end{tabular}}
 \\
 
 * These methods required a combination of multiple imputation and bootstrapping to deal with missing data and derive corresponding standard errors.
\end{minipage}

\end{table*}

\section{Discussion}
\label{s:discussion}

In this paper, we have described different treatment effect estimators and their implementations to account for the use of rescue medication and treatment discontinuation through the hypothetical strategy in a diabetes trial. Overall, the different estimators yielded similar results even though they were based on partially different sets of assumptions (both structural and modelling) and used different information. Even though for this trial IPTW yielded the smallest treatment effect compared to the other methods, there is no reason to expect a similar pattern in other datasets. The impact of the choice of estimator is likely to increase with increasing prevalence of the ICE and missing data, which were both relatively low in the trial we analysed.  

The different estimators considered here can be applied using standard software. As we described in the methods section, there are different R packages available for this purpose. The package for IPTW is very flexible but cannot accommodate missing values. This limitation of handling missing intermittent values can be overcome with multiple imputation combined with bootstrapping which may be computationally intensive. As already explained the package for G-formula allows to use datasets with missing values but still requires  bootstrapping. An attractive alternative is G-formula via MI that avoids the bootstrapping step and is much faster. We recently developed a package to facilitate its implementation \citep{Bartlett2023GviaMI}.

For methods whose implementation in R use the dataset in wide format (MI, G-formula via MI and G-estimation), separate models are fitted at each visit. These models are by default more flexible than those using the long form dataset (IPTW and G-formula) because they allow for different covariate effects and intercept at each visit. The long form dataset implementations use a more parsimonious model by default which has the advantage of improved precision but at the expense of potential bias due to model misspecification. Thus, more complex model specifications are required to achieve the same flexibility as the wide format implementations. For example, in the G-formula model for HbA1c, we included randomised treatment and earlier values of HbA1c (by including the value at the previous visit and the average until the visit) as covariates. An alternative would be to additionally include time or visit as a categorical variable and its interaction with treatment to more flexibly model the evolution of the treatment effects. 

When choosing between estimators, the clinical context has to be taken into account. In some cases, it may be sensible to use information after the ICE occurrence. An advantage of exploiting post-ICE values is a potential gain in efficiency. Thus, methods discarding these data are less powerful than those that use post-ICE data, although in the present analysis this was not evident. However, using post-ICE values comes at the expense of additional modelling assumptions, as we must model the effect of the ICE occurrence on future outcomes. Among those estimators that use only pre-ICE data, IPTW is expected to be less efficient than MMRM/MI \citep{Carpenter2006Comparison}. 

When choosing an estimator, we can also consider accounting for each ICE separately. For example, patients may discontinue treatment due to toxicity whereas they may need rescue medication when their glucose levels are above the expected threshold. These mechanisms can be modelled separately using IPTW allowing for the impact of the covariates on the occurrence of each ICE to differ. Whereas in G-formula, G-formula via MI and G-estimation the effect of an ICE on subsequent outcomes can be allowed to differ according to the type of ICE that has occurred. However, if there are few events of each type of ICE, modelling them separately may become infeasible. It is worth noting that accounting for different ICE mechanisms is only possible with the methods that exploit post-ICE information. 


Table \ref{tab:guidance} provides a summary of the different considerations we have discussed that could help to decide between the different estimators. This could serve as guidance to choose the most appropriate method for a particular setting. 

\begin{table*}
 \centering
 \def\~{\hphantom{0}}
  \caption{Summary of the issues to consider when choosing between estimators}
    \label{tab:guidance}
    \resizebox{\textwidth}{!}{
    \begin{tabular}{ m{5cm} | m{6cm}| m{6cm} }
        Decision & Options & Consideration  \\
        \hline
        Whether to include post-ICE values & Including or not including post-ICE values & Trade off between potential efficiency gain and risk of model mispecification \\
        \hline
        What variables to include & Including additional baseline and time-varying covariates  & Making sequential exchangeability and MAR assumptions more plausible. \\
        \hline
        What modelling approach to use & Using pooled or sequential models & Trade off between efficiency and model flexibility \\
        \hline
        How to handle missing data & Complete case analysis or multiple imputation & MI more efficient. Plausibility of complete case vs MAR missingness assumption \\
        \hline
        How to model ICE occurrence & Single or categorical indicator for each ICE  & Feasibility for modelling separate ICE mechanisms \\
    \end{tabular}}

\end{table*}

In this paper we considered the hypothetical estimand that was targeted by the analysis performed from the original trial publication. 
Different stakeholders may be interested in different estimands. For example, when seeking approval for a novel drug from a regulatory agency, an estimand that `isolates' the effect of the study drug from the rescue medication could be of interest, given that rescue medication cannot be withheld for ethical reasons. In this case, the use of rescue medication would be handled using the hypothetical strategy. Patients and their carers may want to know what they would gain by adhering to the treatment regime, which would translate to an estimand where treatment discontinuation is also dealt with using the hypothetical strategy. On the other hand, health providers or government agencies may be interested in knowing the benefit of the novel drug in routine practice, which includes the potential for using rescue medication and discontinuing the drug. Here, both ICEs would be handled with a treatment policy strategy. However, it is important to note that one is then relying on an assumption that initiation of rescue and discontinuation in the trial is reflective of what would be seen in routine practice. 

In any case, it is arguably unrealistic to conceive of a setting where a patient would continue on treatment in the presence of (serious) adverse events. In such cases, we could consider treatment discontinuation due to adverse events as a different ICE from treatment discontinuation for other reasons, probably unrelated to the treatment. Depending on what happened to patients in the trial after discontinuation of randomised treatment due to adverse events, it may be reasonable to use the treatment policy strategy for such ICEs. If patients were always to receive additional effective medication after discontinuation regardless of the cause, then a hypothetical strategy could still be of interest. 

All of the above considerations regarding choosing the strategy to deal with a particular ICE resemble the considerations of the target trial emulation framework \citep{Hernan2016BigData}. With this framework, observational studies are analysed to estimate the treatment effect in an ideal trial. The idea of target trial emulation was developed to draw causal inference from observational data when RCTs where not ethical, feasible or not yet available. For trials, it may be unethical to randomly assign insulin as rescue therapy for patients with high glucose levels, but it is possible to imagine a hypothetical trial where this could be the case. With such a target trial in mind, the target estimand can be described.

We hope that these considerations regarding the suitability of a hypothetical strategy to handle a particular ICE in a given context, the step-by-step description of different available estimators and further considerations of the implications of their different underlying assumptions are useful for planning, conducting and analysing trials using the estimand framework. 


\section*{Acknowledgements and Conflicts of Interest}

This work was funded by a UK Medical Research Council grant (MR/T023953/1). David Wright is a full time employee of AstraZeneca and owns shares in AstraZeneca and provided some of his time to support Camila Olarte Parra in her research. This publication is based on research using data from data contributors AstraZeneca that has been made available through Vivli, Inc. Vivli has not contributed to or approved, and is not in any way responsible for, the contents of this publication. Data underlying the findings described in this manuscript may be obtained in accordance with AstraZeneca’s data sharing policy described at \url{https://astrazenecagrouptrials.pharmacm.com/ST/Submission/Disclosure}. AstraZeneca Vivli member page is also available outlining further details: 
\url{https://vivli.org/ourmember/astrazeneca/}.

\vspace*{-8pt}


\bibliographystyle{biom} 
\bibliography{mybiblo.bib}

\end{document}